\documentclass[letter]{aa}
\usepackage[varg]{txfonts}
\usepackage{graphicx}
\usepackage{amsmath}
\usepackage{hyperref}
\usepackage{multirow}
\usepackage{lscape}

\DeclareUnicodeCharacter{2003}{}
\DeclareUnicodeCharacter{2212}{}

\begin{document}

\title{Dynamic cosmography of the local Universe: Laniakea and five more watershed superclusters}

\author{A.~Dupuy\thanks{adupuy@kias.re.kr}\inst{1}\and 
H.M.~Courtois\inst{2}
}

\institute{Korea Institute for Advanced Study, 85, Hoegi-ro, Dongdaemun-gu, Seoul 02455, Republic of Korea
\and Universit\'e Claude Bernard Lyon 1, IUF, IP2I Lyon, 4 rue Enrico Fermi, 69622 Villeurbanne, France
} 

\date{Received A\&A March, 15, 2023 - AA/2023/xxxxx; Accepted date}

\abstract {

This article delivers the dynamical cosmography of the Local Universe within z=0.1 (1 giga light-years). We exploit the gravitational velocity field computed using the CosmicFlows-4 catalog of galaxy distances to delineate superclusters as watersheds, publishing for the first time their size, shape, main streams of matter and the location of their central attractor. Laniakea, our home supercluster's size is confirmed to be 2 $\times 10^6$ (Mpc $h^{-1}$)$^3$. Five more known superclusters are now dynamically defined in the same way: Apus, Hercules, Lepus,  Perseus-Pisces and Shapley. Also, the central repellers of the Bootes and Sculptor voids are found and the Dipole and Cold Spot repellers now appear as a single gigantic entity. Interestingly the observed superclusters are an order of magnitude larger than the theoretical ones predicted by cosmological $\Lambda$CDM simulations.

}


\titlerunning{Gravitational basins in CF4}
\authorrunning{Dupuy et al.} 
\maketitle

%

\section{Introduction}

Cosmography is the science that maps and measures the large-scale structures in the observed Universe that are built from the tug of war between gravitation and space expansion. Mapping the position and spatial extents of clusters, filaments, walls, superclusters and voids of galaxies is most frequently and more easily done using the Hubble-Lema\^itre law on redshift datasets. However, such positions and sizes are distorted by the local gravitational velocity field that curves clusters of galaxies and elongates them radially to the observer. This local effect also squeezes the thickness of the walls of galaxies that appear more concentrated, less thick in redshift space than they are in real space. Several methodologies are developed to counter these distortions that will be dominant as bigger and bigger datasets arrive.

However, in the local Universe, i.e below $z=0.1$, the cosmography can be studied using direct distance measurements of galaxy positions that are much less affected by redshift space distortion effects, but are numerically much fewer and consequently prone to strong incompleteness biases collected under the name of "Malmquist biases". The gravitational (a.k.a. peculiar) velocity field derived from direct galaxy distances reconstructs the underlying distribution of mass responsible of these motions, hence allowing a dynamic cosmography of the (dark and luminous) matter distribution.

Such a dynamical mapping of the Local Universe volume also provides an evolution in the semantic segmentation of large-scale structures with the possibility to compute watersheds, basins of attraction and basins of repulsion. Static filaments, walls and voids are embedded into a more global view of large dynamical superclusters delineated by empty regions. The main advantage of mapping superclusters as watershed basins in the divergence velocity field is the robustness of the definition. For example, it allows to quantitatively compare the observed sizes to the predicted ones in cosmological simulations \citep{2021MNRAS.500L..32P} enabling superclusters to be used as cosmological probes.

In mirror, in the hierarchy of structures, voids are regions empty of galaxies, that can be described as large coherent volumes, evacuating flows of matter \citep{2013MNRAS.428.3409A,2022arXiv221116388C}.  Dynamic mapping can thus also help using Cosmic voids as probes of cosmology and galaxy evolution models \citep{2007MNRAS.380..551P, 2022JCAP...12..028F, 2023NatureDominguez}.

In this study, large scale structures are defined as regions of space with gravitationally induced coherent inward or outwards motions, which, in this way, describe basins of attraction or basins of repulsion, respectively. In other words, a gravitational basin is the volume of space containing all the motions of mass going towards a common center, defined as an attractor in the case of basins of attraction, or repeller for basins of repulsion. As per this definition, we may also describe basins as \textit{watersheds}. We obtain then two tessellations of the velocity field into gravitational basins: into basins of attraction on one side, and into basins of repulsion on the other side. A point in space belongs to at most one basin of attraction and one basin of repulsion.

This article presents the cosmography of the Local Universe within $z=0.1$ by partitioning the volume of space into gravitational watersheds and delivering the positions of core attractors, main cosmic streams and strong repellers at the center of voids. Section \ref{sec:CF4} presents the computation of the peculiar velocity field using the CosmicFlows-4 catalog of galaxy distances \cite{2023A&A...670L..15C}. The methodology to define superclusters as basins of attraction \cite{2019MNRAS.489L...1D} is explained in section \ref{sec:method}. Section \ref{sec:resultsBOA} delivers the sizes and locations of the newly cartographied local large-scale structures defined as basins of attraction. An short analysis of streamlines and gravitational valleys is presented in Section \ref{sec:hstream}. Structures defined as repellers and basins of repulsion are discussed in Section \ref{sec:resultsBOR}. Conclusions are drawn in Section \ref{sec:conclusion}.

\begin{figure*}[!h]
\begin{center}
\includegraphics[width=\textwidth,angle=-0]{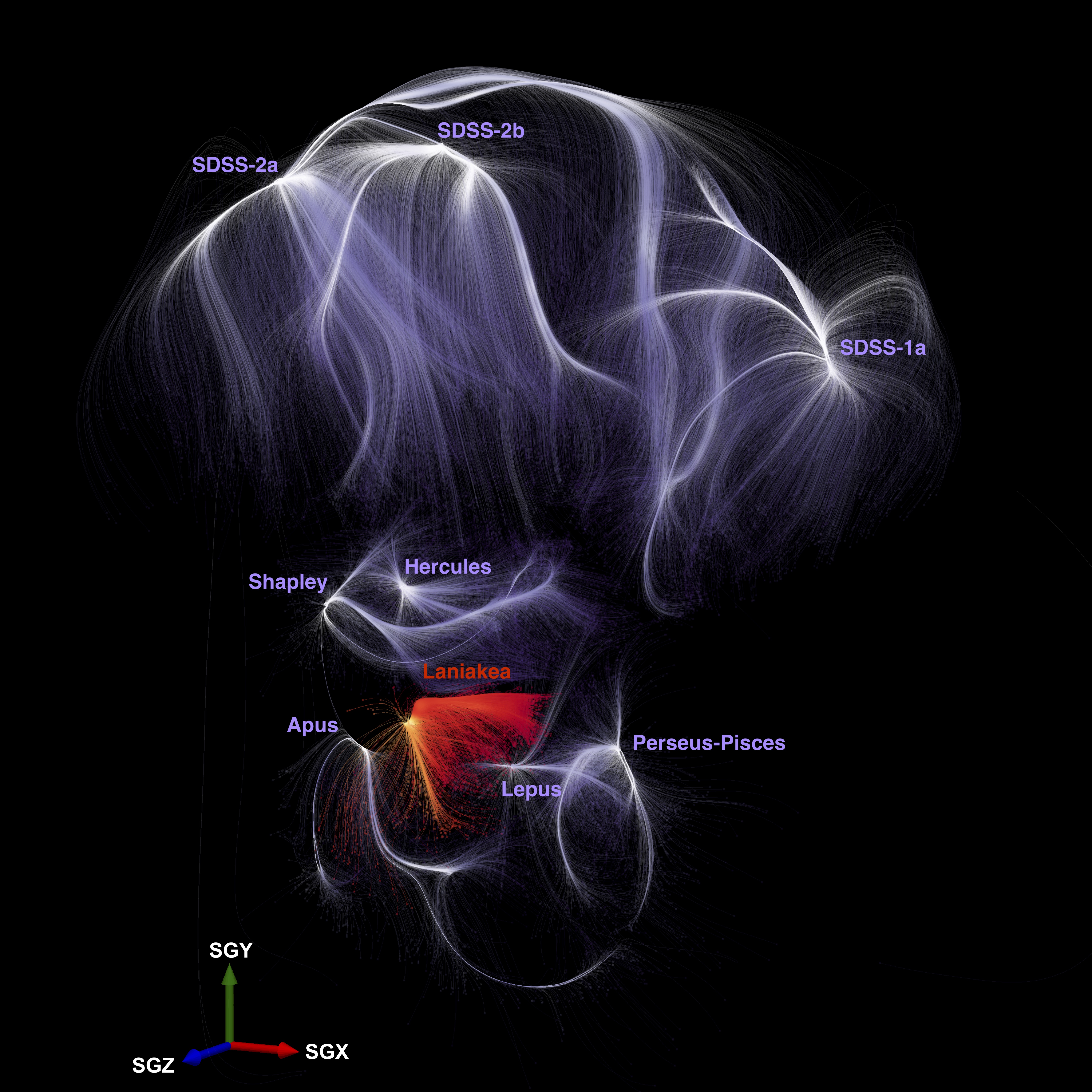}
\caption{Visualization of the basins of attraction of Laniakea (in red) and the five superclusters that are now dynamically defined as watershed in the same way, namely Apus, Hercules, Lepus, Perseus-Pisces and Shapley, as well as the basins identified further away in the SDSS area (both in purple). The basins shown are obtained from the ungrouped CosmicFlows-4 (CF4) velocity field. Each galaxy from the CF4 catalog is positioned at its redshift and represented by tiny dots, red for galaxies part of the Laniakea basin of attraction and purple otherwise. Streamlines are also computed for each galaxy with the same color code. The gradient of color is related to the density of streamlines. Streamlines colored white or yellow indicate a high density of streamlines. Each identified basin of attraction is annotated by the associated structure name. For reference, the three supergalactic cartesian orientation axes SGX, SGY, SGZ are drawn on the bottom left, represented respectively by red, green and blue arrows of size 50 Mpc $h^{-1}$.} 
\label{fig:visuCF4}
\end{center}
\end{figure*}

\section{Methods}

\subsection{Data: 3-dimensional velocity field}
\label{sec:CF4}

The \textit{CosmicFlows-4} catalog (CF4, \citep{2023ApJ...944...94T}) is currently the latest and largest compilation of reliable distances of galaxies and groups of galaxies independent of redshift, obtained through eight methodologies, such as the Tully-Fisher relation for spiral galaxies (TF, \cite{1977A&A....54..661T}), or Fundamental Plane for elliptical galaxies (FP, \cite{1987ApJ...313...42D,1987ApJ...313...59D}). It provides distances for 55,877 galaxies (Table 2 in \cite{2023ApJ...944...94T}), and distances for 38,065 groups constructed from the groups of \cite{2017ApJ...843...16K, 2015AJ....149..171T, 2017A&A...602A.100T} \citep[Table 3 in ][]{2023ApJ...944...94T}, with a uniform coverage of the sky up to approximately 8,000 km s$^{-1}$. Two FP samples provides more distant coverage in two distinct patches of the sky: up to 16,000 km s$^{-1}$ in south celestial hemisphere (6dF Galaxy Survey or 6dFGSv, \cite{2014MNRAS.445.2677S}) and up to 30,000 km s$^{-1}$ in the region covered by Sloan Digital Sky Survey (SDSS, \cite{2000AJ....120.1579Y,2022MNRAS.515..953H}). 

From the CF4 galaxy and group distances, it is possible to reconstruct the three-dimensional overdensity $\delta$ and related peculiar velocity $\vec{v}$ fields, by using the  method introduced in \cite{2023A&A...670L..15C}. The algorithm uses a Hamiltonian-Monte-Carlo (HMC) algorithm which derives the 3D overdensity and velocity fields through exploring a range of free parameters (matter density parameter $\Omega_m$, bias, ...). The reader can refer to \cite{2023A&A...670L..15C} for more details regarding the reconstruction methodology. In this article, all reconstructions have been derived by considering a $\Lambda$CDM cosmology for the initial values: $\Omega_m = 0.3$, and a value of the Hubble constant $H_0 = 74.6$ km s$^{-1}$ Mpc$^{-1}$, compatible with the CF4 assembly of distances as stated in \cite{2023ApJ...944...94T}. In order to reach a stable convergence on the free parameters, about 10,000 HMC steps have been computed and the resulting overdensity and peculiar velocity fields are obtained by averaging all steps (except the burning steps). 

We include in this article two 3D reconstructions of the overdensity and velocity fields: one reconstruction derived from the galaxy sample of the CF4 dataset, and another reconstruction obtained from groups of galaxies also provided by the CF4 compilation. Both reconstructions are computed in a grid of size $128^3$ and side length of $1,000$ Mpc $h^{-1}$, hence reaching a resolution of 7.8 Mpc $h^{-1}$. Figures displaying the overdensity and velocity fields reconstructed from both CF4 galaxies and groups are included in Appendix \ref{appendix}. Figure \ref{fig:Vattractors} shows three-dimensional visualizations of the reconstructed overdensity field through 4 levels of isosurfaces ranging from white to dark red. The reconstruction from CF4 galaxies is shown on the top panel, while the reconstruction from CF4 groups is shown on the bottom panel. The related peculiar velocity field is represented by streamlines. The orientation of the visualization is given by the red, green, and blue arrows of length 50 Mpc $h^{-1}$, directed along the supergalactic cartesian SGX, SGY and SGZ axes respectively and located at the center of the reconstructed cube. More details on this plot are given in Appendix \ref{appendix}. We can recognize known structures in both reconstructions, such as Norma close to the center, Shapley supercluster in -SGX and +SGY, Perseus-Pisces supercluster on the +SGX side, as well as Apus, Pisces-Cetus, or Hercules superclusters. However the two reconstructions differ since the grouped version shows a strong accent on the more distant SDSS region in +SGY. The grouped reconstruction is therefore less satisfying than the ungrouped one, since we notice a "ring" that appears due to a lack of constraints in the ortho-radial direction around the observer and also arising from the intrinsic shaped and size of the SDSS galaxy distances catalog. 

\subsection{Watersheds and gravitational basins}
\label{sec:method}

The methodology we use to identify attractors, repellers and their respective gravitational basins in this article has been introduced in \cite{2019MNRAS.489L...1D}. In the case of a time-independent linear velocity field, streamlines are paths that are always tangent to the local value of the velocity field. They are computed by spatially integrating the components of the peculiar velocity field, starting from a given seed point. The integration is performed by the fourth order Runge-Kutta (RK4) numerical integrator. Streamlines converge towards the critical points of the velocity field, i.e \textit{attractors} or \textit{repellers}. Hence depending on the direction of integration considered, either forward (integrating $\vec{v}$) or backward (integrating $-1 \times \vec{v}$), we can identify the positions of attractors or repellers, respectively, by finding the points of convergence of streamlines. Then, a gravitational basin can be defined as a set of seed points whose streamlines converge towards the same critical point of the velocity field: an attractor for \textit{basins of attraction}, or a repeller for \textit{basins of repulsion}. 

As the peculiar velocity is available as a grid, streamlines are actually computed for each voxel (a cell in three dimensions) of the velocity grid with a fixed number of RK4 iterations. Attractors (or repellers) are then identified as the voxels where most streamlines are terminated, also called \textit{ending points}. The velocity grid is then segmented into basins of attraction (or repulsion) defined as sets of (seed) voxels whose streamlines share the same ending point. As streamlines are computed with a fixed number of iterations, they may be terminated before reaching a critical point of the velocity field. Hence several iterations are needed when matching voxels to basins, making the basins "grow" around the attractors or repellers. Finally, we obtain two tessellations of the velocity field: into basins of attraction and into basins of repulsion, depending on the direction of integration. Each voxel of the velocity field belongs to, at most, one basin of attraction and one basin of repulsion. Since the considered peculiar velocity grid does not have periodic boundary conditions, seed voxels whose streamlines go out of the computational grid do not belong to any basin. More details on the basin segmentation algorithm can be found in \cite{2019MNRAS.489L...1D}. More extensive testing of this methodology has been conducted using cosmological simulations in \cite{2020MNRAS.493.3513D}.

Additionally, the methodology of segmenting basins allows the definition of a new quantity that helps visualizing gravitational valleys. Instead of analyzing the convergence of streamlines to tessellate the velocity field into gravitational basins, one can simply construct an histogram from the streamlines by counting how many streamlines are passing through each voxel, and assign that value to the associated voxel to build the histogram. This cube, of the same size and resolution as the grid containing the velocity field, produces a map of the density of streamlines, with regions high or low in streamlines concentrations, showcasing gravitational valleys. This definition allowed a kinematic confirmation of Vela supercluster, known to be hidden within the Zone of Avoidance \citep{2019MNRAS.490L..57C}. We also refer the reader to \cite{2019MNRAS.489L...1D} to find more details about that algorithm.


\section{Watersheds of the Local Universe and their core attractors}
\label{sec:resultsBOA}

The methodology of segmenting watersheds of basins of attraction was performed on both versions of the CosmicFlows-4 (CF4) catalog (ungrouped and grouped galaxies). The main body of the article presents the most robust analysis obtained using individual galaxy distances (ungrouped CF4). However all results for the grouped version are available in detail in Appendix \ref{appendix}. Nine basins of attraction were found in the velocity field reconstructed from the CF4 individual galaxies. They can be visualized in Figure \ref{fig:visuCF4}. The new basin of attraction of our home supercluster Laniakea is shown in red, other basins, namely the five superclusters now dynamically defined as watersheds (Apus, Hercules, Lepus, Perseus-Pisces and Shapley) and basins identified in the area covered by SDSS, are shown in purple. Streamlines are computed from each individual CF4 galaxy, which are themselves represented by tiny dots. The gradient of color (from red to yellow and purple to white for Laniakea and other basins, respectively) is related to the density of streamlines. White or yellow streamlines denote a high concentration of streamlines in that area. Each basin of attraction is annotated by the name of the associated supercluster. For practical purposes, the three supergalactic cartesian orientation axes SGX, SGY and SGZ are displayed at the bottom left of the Figure, rendered by red, green and blue arrows of size 50 Mpc $h^{-1}$, respectively.



The basin of attraction corresponding to the Laniakea supercluster is only identified in the ungrouped reconstruction. In the grouped reconstruction it appears as part of the Shapley basin of attraction. The segmented Laniakea basin has a volume of $1.9 \times 10^6$ (Mpc $h^{-1}$)$^3$. It can be compared to the earliest estimates of CosmicFlows-2 (CF2) dataset: eye-ball measurement computed a decade ago giving $1.7 \times 10^6$ (Mpc $h^{-1}$)$^3$ in \cite{2014Natur.513...71T} and using the segmentation algorithm on CF2 data gives $2.3 \times 10^6$ (Mpc $h^{-1}$)$^3$ .
The Laniakea central attractor is located at $[-62, -8, 39]$ Mpc $h^{-1}$ in supergalactic cartesian coordinates, equivalent to $sgl=187^\circ$, $sgb=32^\circ$ and $cz=7370$ km s$^{-1}$. This is in full accordance with the direction of the Great Attractor \citep{1987ApJ...313L..37D}. 


In the same way as Laniakea, the Apus supercluster is only identified as a single basin of attraction in the velocity field reconstructed from individual CF4 galaxies. Its associated attractor is found to be located at $[-125,-23,-31]$ Mpc $h^{-1}$, while the segmented basin has a volume of $9.5 \times 10^6$ (Mpc $h^{-1}$)$^3$. 

The attractor coordinates for the Hercules supercluster show a great coherence between the two different reconstructions, being detected at $[-39,70,78]$ Mpc $h^{-1}$ and $[-47,78,78]$ Mpc $h^{-1}$ for ungrouped and grouped galaxies, respectively, although the volumes of the basin differ a lot ($3.1 \times 10^6$ (Mpc $h^{-1}$)$^3$ for individual galaxies and $0.8 \times 10^6$ (Mpc $h^{-1}$)$^3$ for galaxy groups), mainly due to stronger emphasis of the SDSS data in the grouped CF4. We can compare the Hercules attractor coordinates to the position of the Abell cluster A2151 \citep{1989ApJS...70....1A}, also known as the Hercules cluster of galaxies, with a right ascension and declination of R.A. = $241^{\circ}$ and Dec. = $+17^{\circ}$.

The Lepus supercluster is a basin of attraction in both reconstructions, with a volume of $8.1 \times 10^6$ (Mpc $h^{-1}$)$^3$ for individual galaxies and $6.9 \times 10^6$ (Mpc $h^{-1}$)$^3$ for galaxy groups, around an attractor positioned at $[-62,-39,-109]$ Mpc $h^{-1}$ and $[-47,-55,-125]$ Mpc $h^{-1}$, respectively. 

The Perseus-Pisces supercluster basin of attraction displays a volume of  $4.8 \times 10^6$ (Mpc $h^{-1}$)$^3$ in the individual galaxies reconstruction, and $2.0 \times 10^6$ (Mpc $h^{-1}$)$^3$ in the reconstruction obtained from CF4 galaxy groups. Its central attractor is located at $[47,-23,-31]$ Mpc $h^{-1}$ and $[47,-23,-39]$ Mpc $h^{-1}$ for CF4 galaxies and groups, respectively, showing an important consistency between the two reconstructions. The coordinates of the Perseus-Pisces attractor differ from the location of the richest cluster in the Perseus-Pisces supercluster, the Abell cluster A426, with R.A. = $50^{\circ}$ and Dec. = $+41^{\circ}$ \citep[also see][]{1978MNRAS.185..357J, 2015AJ....149..171T}. However due to the low resolution of the reconstructions considered in this manuscript (7.8 Mpc $h^{-1}$), the positions derived from the segmentation method are not accurate and cannot be compared to redshift positions of clusters.

Last, the Shapley supercluster has also been identified as a basin of attraction around an attractor located at $[-141,62,-16]$ Mpc $h^{-1}$ in both velocity fields. The position of the attractor is in good agreement with, for example, two Abell clusters known to be part of the Shapley supercluster: A3570 located at R.A, Dec. = $(207^{\circ}, -37^{\circ})$, and A3575 at R.A, Dec. = $(208^{\circ}, -32^{\circ})$. The volume of Shapley is evaluated to be $7.9 \times 10^6$ (Mpc $h^{-1}$)$^3$ and $27.0 \times 10^6$ (Mpc $h^{-1}$)$^3$ for CF4 ungrouped and grouped galaxies, respectively. In both cases, the basin includes the Coma supercluster. This may be an indication that Coma supercluster could be a sub-basin of Shapley supercluster. Shapley's basin found using the CF4 grouped version is much larger as it also encompasses the Apus and Laniakea superclusters, each identified as separate basins in the reconstruction from CF4 ungrouped galaxies.

A few attractors have also been identified in the more distant region covered by the SDSS addition to the CF4 catalog. However, the computed basins of attraction cannot be constrained correctly because streamlines are overflowing up to the borders of the computational box, due to the lack of further distant data needed to delimit the basins on their far side. Also the errors on the galaxy distance data are proportional to distance, for this reason we will not comment on the volumes of the basins, though the instability in the values between the two reconstructions is easy to notice. For the sake of clarity, the SDSS attractors are classified in two groups depending on their position in the SDSS reconstructed overdensity. In the ungrouped reconstruction, only three attractors are identified. They are labelled \textit{SDSS-1a}, located at $[117,227,-227]$ Mpc $h^{-1}$,
\textit{SDSS-2a/b}, identified accordingly at $[-102,297,102]$ Mpc $h^{-1}$ and $[31,289,156]$ Mpc $h^{-1}$.

The main source of error in the current cosmography is the difference between the two CosmicFlows-4 catalog's variations: individual galaxies and groups of galaxies. The fact of grouping or not galaxy distances impacts strongly the resulting peculiar velocity fields. More work is needed in the preparation of catalogs of galaxy distance in order to mitigate that problem. The grouped version of CF4 gives too much importance to the SDSS subset of data, inflating the signal in that more distant region where the errors in the data are harder to constrain. This results in an overestimation of the SDSS basins sizes.

\section{Streamlines and gravitational valleys}
\label{sec:hstream}

\begin{figure*}[]
\begin{center}
\includegraphics[height=0.26\textheight,angle=-0]{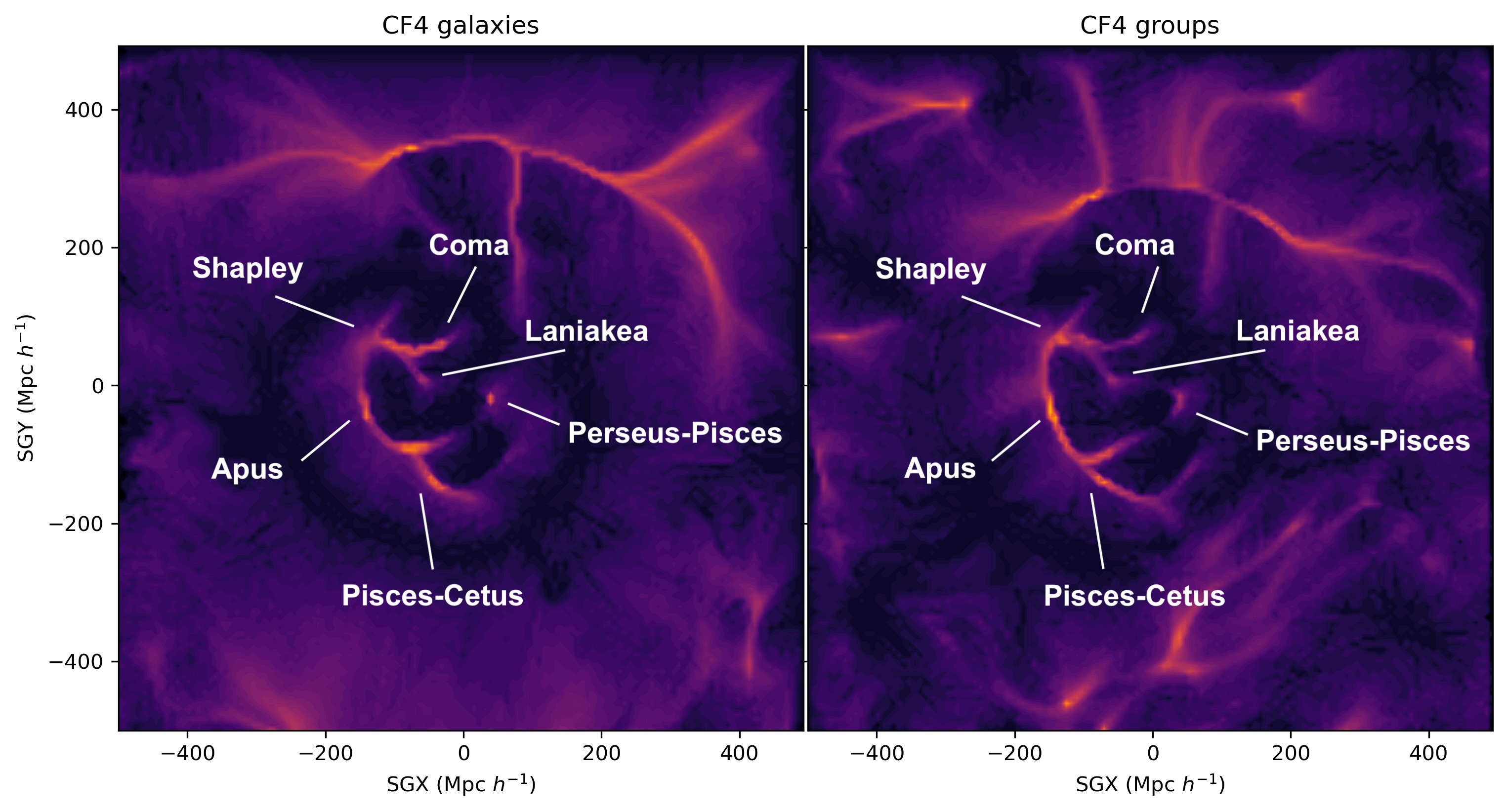} \includegraphics[height=0.26\textheight,angle=-0]{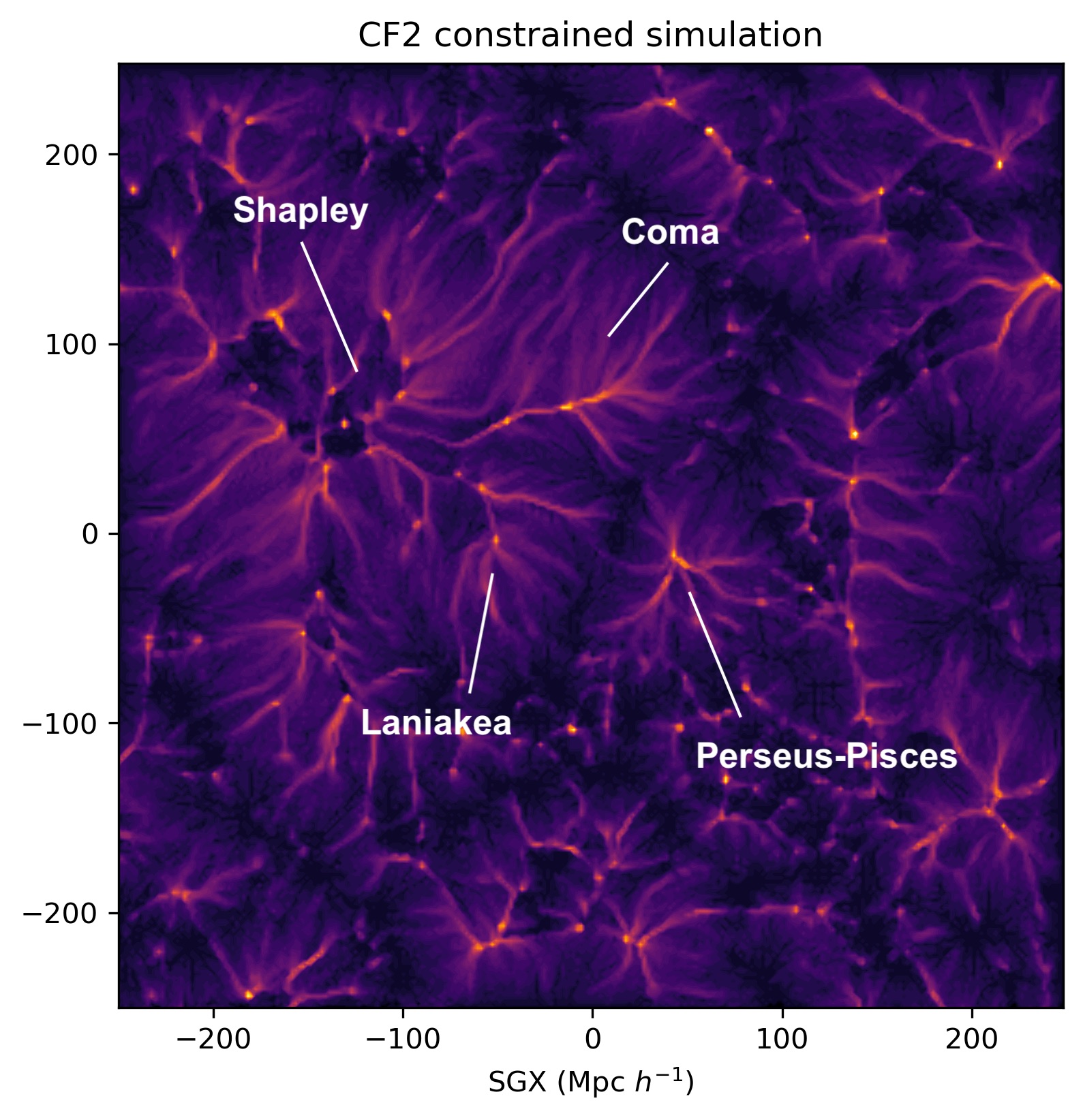}
\caption{Left and middle: Gravitational valleys in the Local Universe. Both panels show a supergalactic cartesian SGX-SGY slice of width 7.8 Mpc/h, centered on SGZ = 0 Mpc $h^{-1}$, of the density of streamlines, derived from the ungrouped (left) and grouped (right) CosmicFlows-4 velocity field. Yellow represents voxels highly concentrated in streamlines, while black represents a low concentration of streamlines. Right: Density of streamlines from a $\Lambda$CDM simulation constrained with the CF2 data, with the same color code. Structures are annotated on the Figure. Note that the scale of the axes is different from the two other panels. As a matter of fact CF2 catalog was much sparser in data (seven times less data) and much smaller in cartographied volume (ten times smaller compared to CF4), thus this simulation is loosely constrained on scales up to 150 Mpc $h^{-1}$ maximum. Also at that time, we had poor coverage of the southern hemisphere large scale structures.}
\label{fig:hstream}
\end{center}
\end{figure*}


Figure \ref{fig:hstream} left and middle panels show the density of streamlines computed from the peculiar velocity field reconstructed from CF4 individual galaxies (left) and groups of galaxies (right). Both panels correspond to a supergalactic cartesian SGX-SGY slice, centered on SGZ = 0 Mpc $h^{-1}$. Regions with a high concentration of streams are shown in yellow, while regions with less streams are shown in black. One can notice how the density of streamlines is depicting a "skeleton" of the peculiar velocity field, as one can spot the superclusters present in this slice. The gradient of yellow and purple present around each structure also helps visually identify the coverage of the associated basins of attraction as described earlier in the manuscript.

\section{Repellers of the Local Universe}
\label{sec:resultsBOR}

Repellers, and associated basins of repulsion, can be identified in a mirroring way as attractors and basins of attraction, by simply computing streamlines in the backward direction, i.e integrating $-1 \times \vec{v}$. The segmenting algorithm detected three repellers (with their associated basin of repulsion) in the ungrouped reconstruction.

A basin of repulsion is identified in the Sculptor Void region, centered on a repeller located at $[23,-109,-16]$ Mpc $h^{-1}$ and encompassing a volume of 1.3 $\times 10^6$ (Mpc $h^{-1}$)$^3$.
Similarly, a repeller in the SDSS area is detected near the Bootes Void region, at $[-31,125,8]$ Mpc $h^{-1}$.
A huge basin of repulsion is spotted at $[133,-133,62]$ Mpc $h^{-1}$ (or $sgl=315^\circ$, $sgb=18^\circ$ and $cz=19,800$ km s$^{-1}$).
This direction is about $20^\circ$ away from the direction of the Dipole Repeller at $sgl\sim334^\circ$, $sgb\sim39^\circ$ and $cz=14,000$ km s$^{-1}$, and $50^\circ$ away from the direction of the Cold Spot Repeller at $sgl\sim287^\circ$, $sgb\sim-16^\circ$ and $cz=23,000$ km s$^{-1}$ \citep{2017NatAs...1E..36H,2017ApJ...847L...6C}.
The Dipole and Cold Spot repellers now appear as a single gigantic extended entity, of volume 163 $\times 10^6$ (Mpc $h^{-1}$)$^3$. 

In the grouped reconstruction, two smaller new regions evacuating matter are cartographied : one in the zone of avoidance and one within the SDSS surveyed volume. A full table is available in Appendix \ref{appendix}, accompanied by three-dimensional visualizations of the detected repellers and their related basins of repulsion.  



\section{Conclusion}
\label{sec:conclusion}

This article presents the dynamic cosmography of the local Universe, up to $z=0.1$, by analysing the gravitational velocity field reconstructed from the Cosmicflows-4 galaxy distances. This analysis allows to identify the large scale structures of the Universe as gravitational basins. A decade after its discovery, we confirm the shape, size  and the coordinates of the attractor of our home supercluster Laniakea, which now englobes a volume of $1.9 \times 10^6$ (Mpc $h^{-1}$)$^3$. Five known superclusters, namely Apus, Hercules, Lepus, Perseus-Pisces and Shapley, are now dynamically defined as basins of attraction or watersheds for the first time, as well as basins detected in the region covered by SDSS. We notice that the supercluster Coma is part of the basin of Shapley, hinting that it may be sub-structure of Shapley. Superclusters and gravitational basins can also be visualized as gravitational valleys, such as in Figure \ref{fig:hstream}. The coverage of CF4 also allows to define a few basins of repulsion, namely the basins of the Sculptor Void, a basin near the Bootes Void, in the region covered by SDSS and in the Zone of Avoidance. The Dipole Repeller and the Cold Spot Repeller are now defined as a single gigantic entity.


It is worth pointing that the observed superclusters defined in this article are larger than the ones found in cosmological $\Lambda$CDM simulations. \cite{2020MNRAS.493.3513D} studies of basins of attraction detected in a constrained cosmological simulation, using the same segmentation algorithm as in this manuscript. A SGX-SGY slice (centered on SGZ = 0 Mpc $h^{-1}$) of the density of streamlines derived from this simulation and extracted from \cite{2020MNRAS.493.3513D} is shown on the right panel of Figure \ref{fig:hstream}, with the same color code. The authors showed that the typical size of basins of attraction is within $1 \times 10^5$ and $1 \times 10^6$ (Mpc $h^{-1}$)$^3$. The simulation being constrained by the Cosmicflows-2 dataset, they defined the simulated basins of Laniakea and Perseus-Pisces, containing a volume of $5 \times 10^5$ (Mpc $h^{-1}$)$^3$ and $7 \times 10^5$ (Mpc $h^{-1}$)$^3$. This difference in volume between the simulated and observed volumes may come from the fact we are comparing structures from a simulation constrained with the Cosmicflows-2 (CF2) data to structures detected in reconstructions from CF4, with a now much larger coverage and larger number of galaxies than CF2.




Errors on the definition of large-scale structures as gravitational basins can be derived in this work from the difference between the two reconstructions of the local velocity field considered, namely from distances of galaxies and of groups of galaxies, as grouping galaxies and their distances has a direct impact on the resulting peculiar velocity field. Particularly, the
grouped version of CF4 gives too much importance to the SDSS
subset of data, overestimating the SDSS basins sizes.

Additionally, identifying attractors of the gravitational watersheds of the local Universe as known rich galaxy clusters may define an interesting follow-up study of this work. For example, the attractor of the SDSS-2c basin of attraction may point to the very rich A2142 Abell cluster, described in \cite{2020A&A...641A.172E}. However as mentioned above, the current resolution of the reconstruction may be too low to get accurate attractor positions, comparable to redshift positions of rich clusters.

At this point in time, the main contingency regarding the identification of large-scale structures comes from the inclusion of the SDSS data into the CosmicFlows catalog which is a composite of several sources of observations. There is great hope that large independent surveys upcoming in the next few years such as WALLABY (Widefield ASKAP L-band Legacy All-sky Blind surveY \cite{2020Ap&SS.365..118K}), bringing 90,000 galaxies up to $z=0.1$, DESI (Dark Energy Spectroscopic Instrument \cite{2016arXiv161100036D}) and 4HS (4MOST Hemisphere Survey \cite{2019Msngr.175....3D}), each deliverying 500,000 galaxies up to $z=0.15$, will further enrich the cosmography of our Universe.

The main part of the article discusses only two figures, however a more detailed description is given in Appendix \ref{appendix}, along with additional figures, a table summarizing all detected basins of attraction and repulsion, and links to interactive three-dimensional visualizations.

\begin{acknowledgements}
HC is grateful to the Institut Universitaire de France for its huge support which enabled this research. HC also acknowledge support from the CNES. AD is supported by a KIAS Individual Grant (PG087201) at Korea Institute for Advanced Study.

\end{acknowledgements}

\section*{Declarations}

\begin{itemize}

\item Materials and Correspondence: 
A. Dupuy is the author to whom correspondence and material requests should be addressed.

\item Competing interests : The authors declare no competing interests.

\item Availability of data and materials : 
The datasets generated during and/or analysed during the current study are available online on the CosmicFlows project page \footnote{\url{https://projets.ip2i.in2p3.fr/cosmicflows/}}. Grids are of the same size as the velocity field discussed in this manuscript (1000 Mpc $h^{-1}$ and $128^3$) and are filled with integers from 0 to $n$ where $n$ is the number of basins detected. Each integer labels a single basin (in the same order as listed in Table \ref{tab:basins}, ignoring blank lines), and the value of all voxels part of a given basin is set to that same integer. Voxels set to 0 are not part of any basin. For example, in the case of basins of attraction in the CF4 galaxy velocity field: voxels filled with 1 are part of Laniakea, 2 of Apus, 3 of Hercules, etc.

\end{itemize}

\bibliography{biblio}{}
\bibliographystyle{aa}

\appendix

\section{Attractors and Repellers}
\label{appendix}

\begin{figure}[h!]
\begin{center}
\includegraphics[width=0.99\columnwidth,angle=-0]{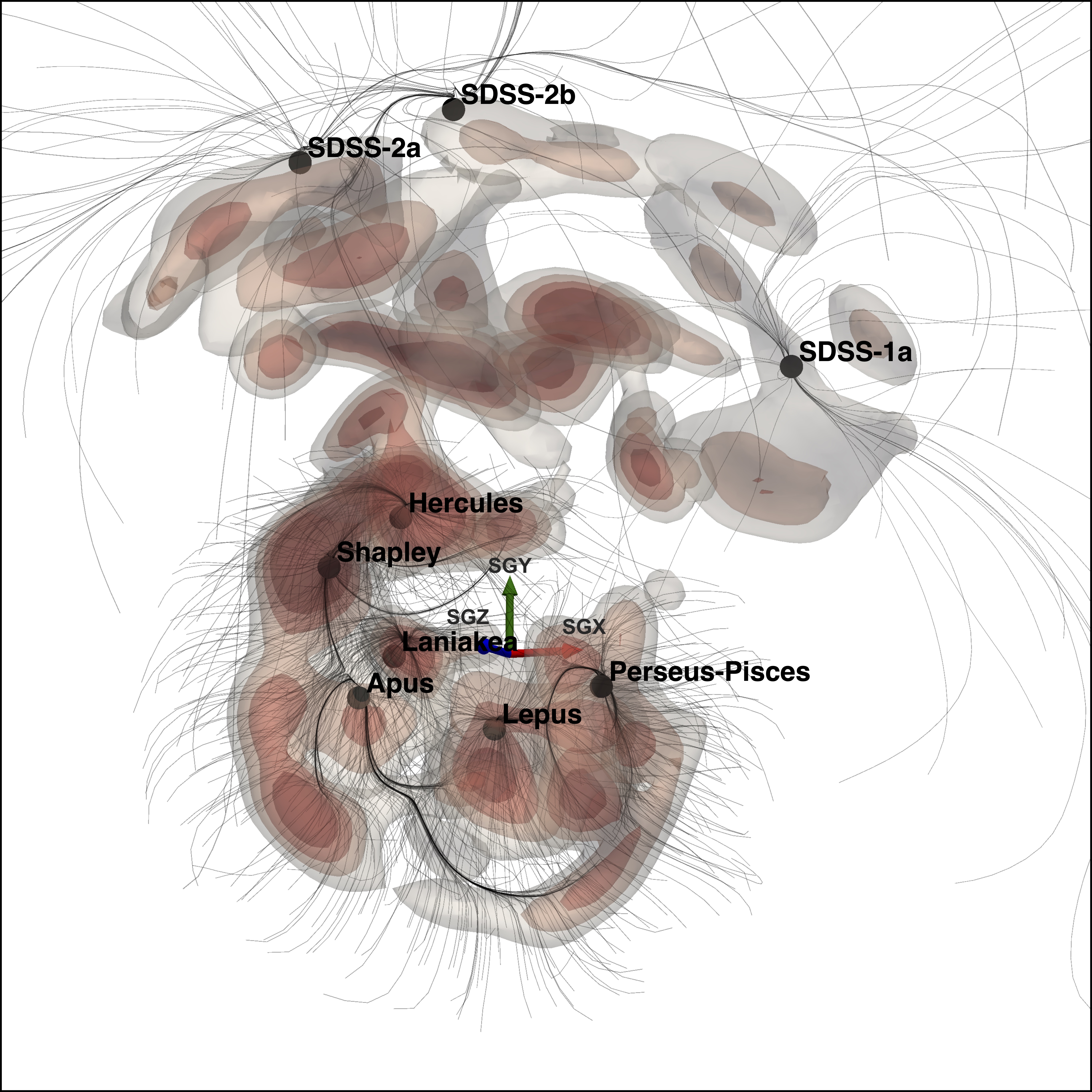}\\
\includegraphics[width=0.99\columnwidth,angle=-0]{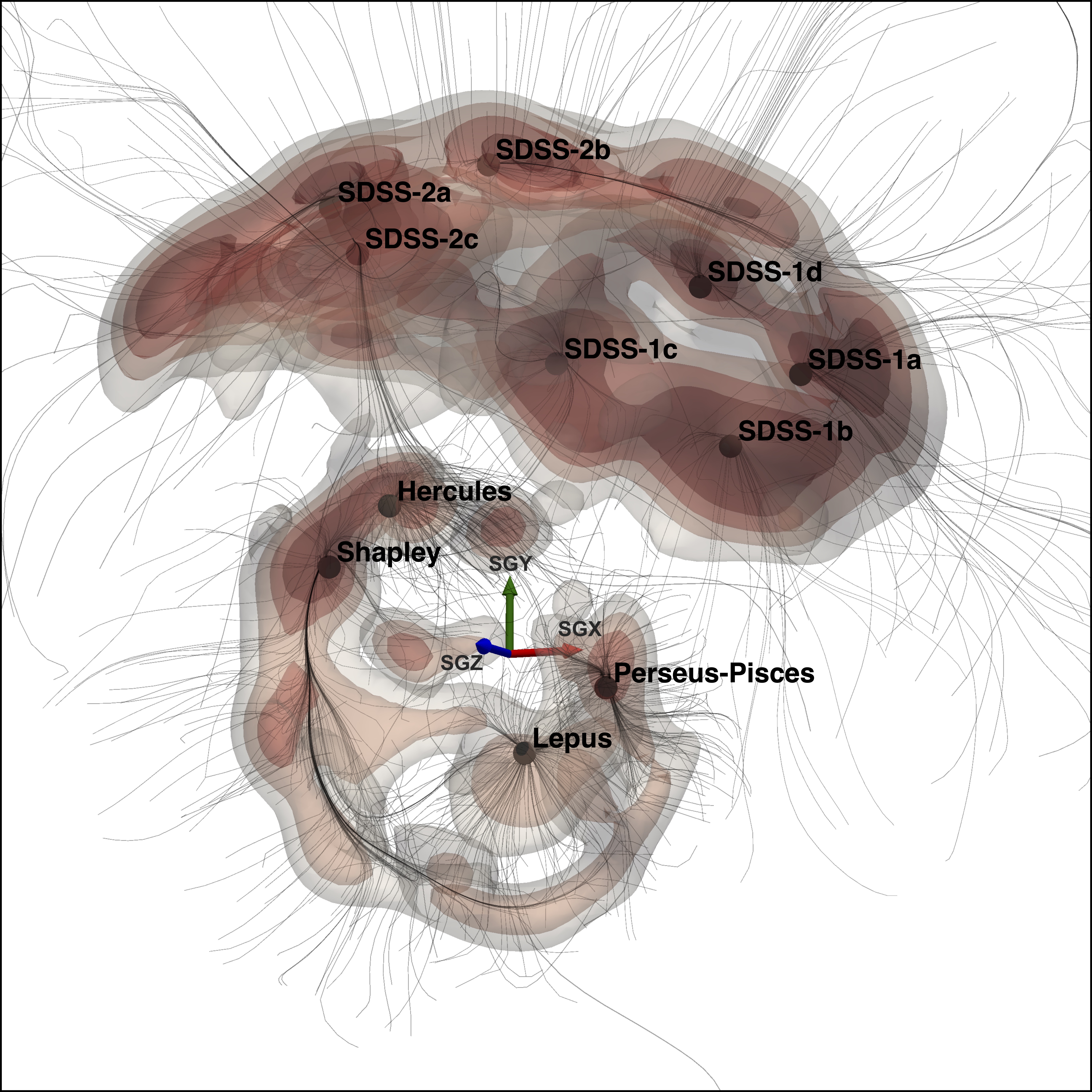}
\caption{Attractors in the local universe reconstructed from the CosmicFlows-4 galaxies (top) and groups (bottom). The reconstructed overdensity field is represented by 4 levels of isosurfaces from white to dark red: 0.75, 1, 1.4 and 1.75 for CF4 galaxies, and 0.5, 0.8, 1.2 and 1.5 for CF4 groups. Streamlines are derived by integrating the reconstructed peculiar velocity field in the forward direction. For clarity purposes, extra noise has been removed by showing only streamlines part of basins of attraction identified with the segmentation algorithm, listed in the top part of Table \ref{tab:basins}. Black spheres are located at the coordinates of attractors identified in the velocity field, i.e at the points where the streamlines are converging. The center of the reconstructed cube is located by the red, green and blue arrows of length 50 Mpc $h^{-1}$ and directing along the supergalactic cartesian SGX, SGY and SGZ axes respectively. Sketchfab interactive visualizations are available for both analysis: \href{https://sketchfab.com/3d-models/attractors-in-reconstruction-from-cf4-galaxies-fb0c18e7930c4b6b94e3f6a30202bd6a}{galaxies} and \href{https://sketchfab.com/3d-models/attractors-in-reconstruction-from-cf4-groups-fa105e2bafba4b949c8b02c463d6b1b2}{groups}.}
\label{fig:Vattractors}
\end{center}
\end{figure}

\begin{figure*}[h!]
\begin{center}
\includegraphics[width=\textwidth,angle=-0]{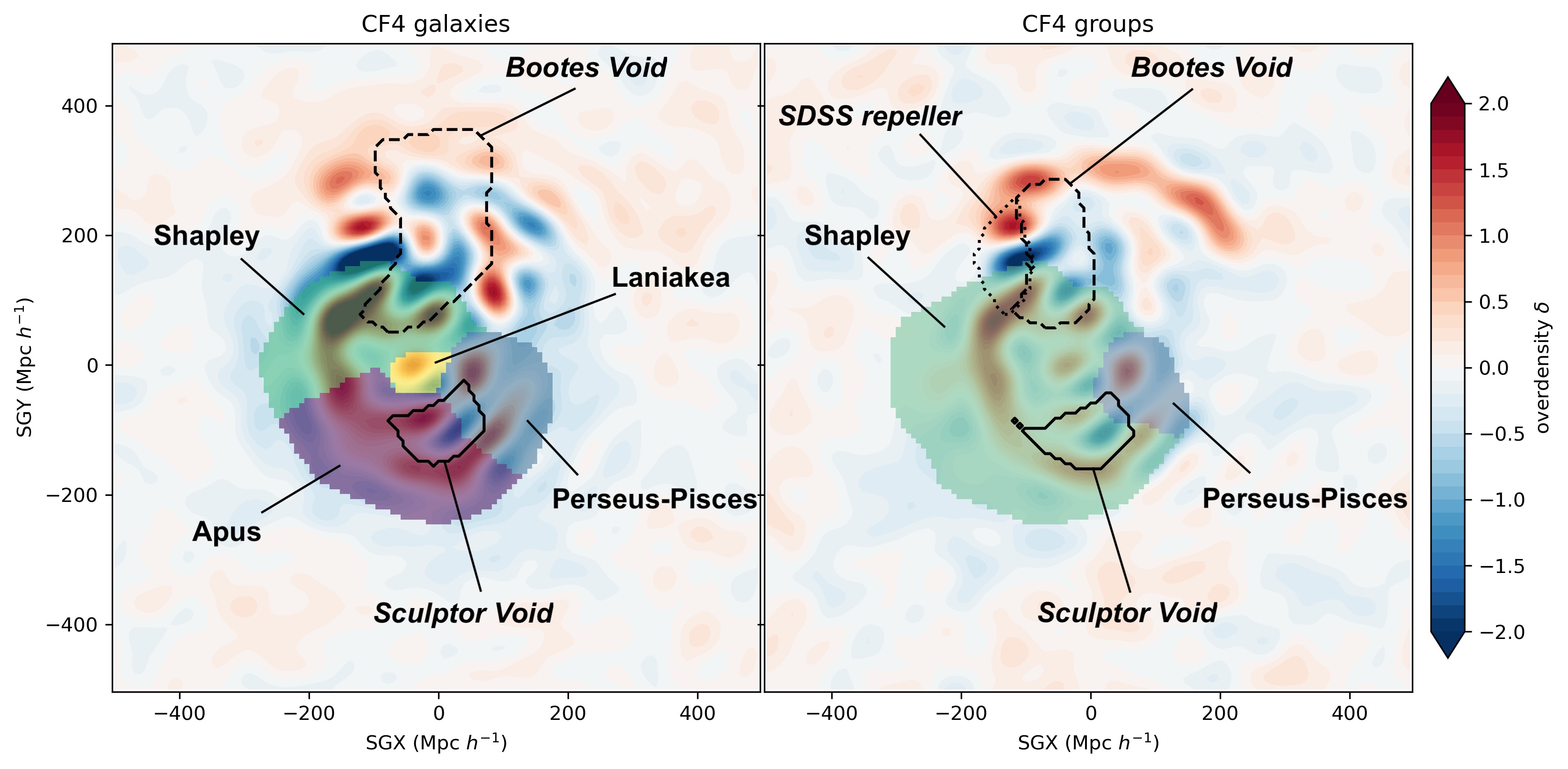}
\caption{SGX-SGY slice, centered on SGZ = 0 Mpc $h^{-1}$ and of width 7.8 Mpc $h^{-1}$, of the overdensity field reconstructed from CF4 galaxies (left) and groups (right), overlapped with the gravitational basins identified in the associated velocity field: basins of attraction (filled contours with no border line) and basins of repulsion (empty contours with solid/dashed/dotted black lines), along with the name of each corresponding structure.}
\label{fig:slice}
\end{center}
\end{figure*}

The methodology described above has been applied to both reconstructions obtained from CF4 individual galaxies and CF4 groups. Structures identified as basins of attraction and repulsion in each reconstruction are listed in the first column of Table \ref{tab:basins} along with their volume in units of 10$^6$ (Mpc $h^{-1}$)$^3$ in columns 5 and 9, and with the supergalactic cartesian coordinates of the respective attractors and repellers in Mpc $h^{-1}$, in columns 2 to 4 and 6 to 8. 
The respective attractors are plotted as black spheres in Figure \ref{fig:Vattractors} with their associated basin of attraction, annotated with the name of the corresponding structure. The reconstruction from CF4 galaxies is shown on the top panel, and the reconstruction from CF4 groups is shown on the bottom panel. The orientation of the two visualizations is given by the red, green, and blue arrows of length 50 Mpc $h^{-1}$, directed along the supergalactic cartesian SGX, SGY and SGZ axes respectively and located at the center of the reconstructed cube. We recall the reader that the 4 levels of isosurfaces ranging from white to dark red are displaying the reconstructed overdensity field, while streamlines represent the related peculiar velocity field. Additionally, for a sense of clarity, streamlines are computed by considering only voxels that are part of all basins of attraction as seed points, allowing to focus on basins identified in the velocity field and remove extra noise caused by streamlines going out of the computational box (hence not part of any basin). For each basin of attraction, the streamlines are nicely converging towards their respective core attractor. Additionally, Figure \ref{fig:slice} shows a SGX-SGY slice, centered on SGZ = 0 Mpc $h^{-1}$, of the overdensity field associated to the velocity field reconstructed from CF4 distances of galaxies (left) and groups (right). The overlapping contours represent the gravitational basins, obtained from the tessellation of the velocity field into basins of attraction (filled contours without borders), and basins of repulsion (empty contours with borders), along with the names of each associated structure.

\begin{figure}[h!]
\begin{center}
\includegraphics[width=0.96\columnwidth,angle=-0]{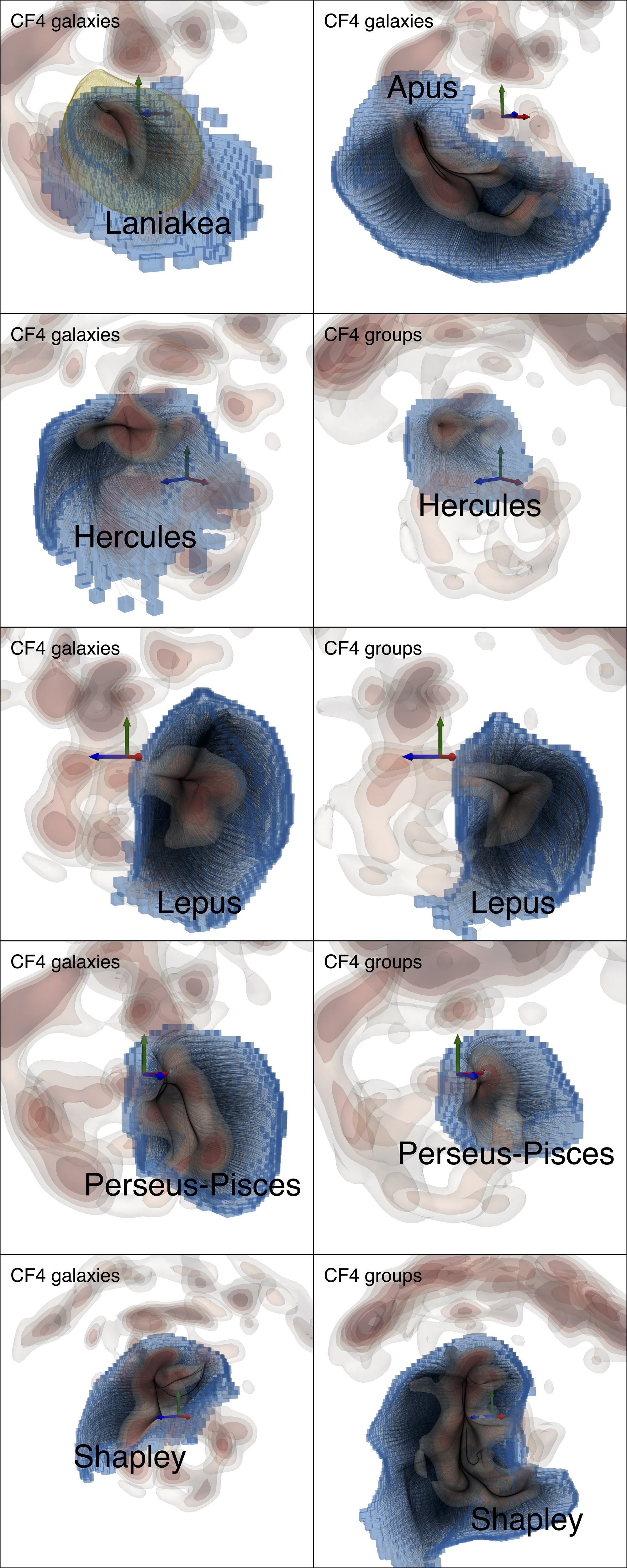}
\caption{Gravitational watersheds identified in the peculiar velocity field reconstructed from CF4 galaxies and groups, each represented by a blue contour and streamlines. In the case of Laniakea, the basin of attraction found in the CosmicFlows-2 reconstruction is also shown in yellow. The reconstructed overdensity field is represented by 4 levels of isosurfaces. The center of the reconstruction is located by the red, green and blue arrows of length 50 Mpc h$^{-1}$ each directing along SGX, SGY and SGZ.}
\label{fig:BOA}
\end{center}
\end{figure}

Figure \ref{fig:BOA} shows zoom-in visualizations of the basins of attraction of six different superclusters: Laniakea, Apus, Hercules, Lepus, Perseus-Pisces and Shapley. The left and right columns are showing basins obtained from the ungrouped and grouped reconstructions respectively, except for the first row (ungrouped version only in this case). In each panel, the overdensity field is represented by four levels of isosurfaces (see Figure \ref{fig:Vattractors} for more details). Streamlines are also computed from the velocity field by considering only voxels part of the basin as seed points. The volume of each basin is shown in blue. In the case of Laniakea, we also show in yellow the volume obtained previously from the peculiar velocity field reconstructed from the CosmicFlows-2 catalog \citep{2019MNRAS.489L...1D} which allowed its discovery. 

\begin{figure}[h!]
\begin{center}
\includegraphics[width=0.99\columnwidth,angle=-0]{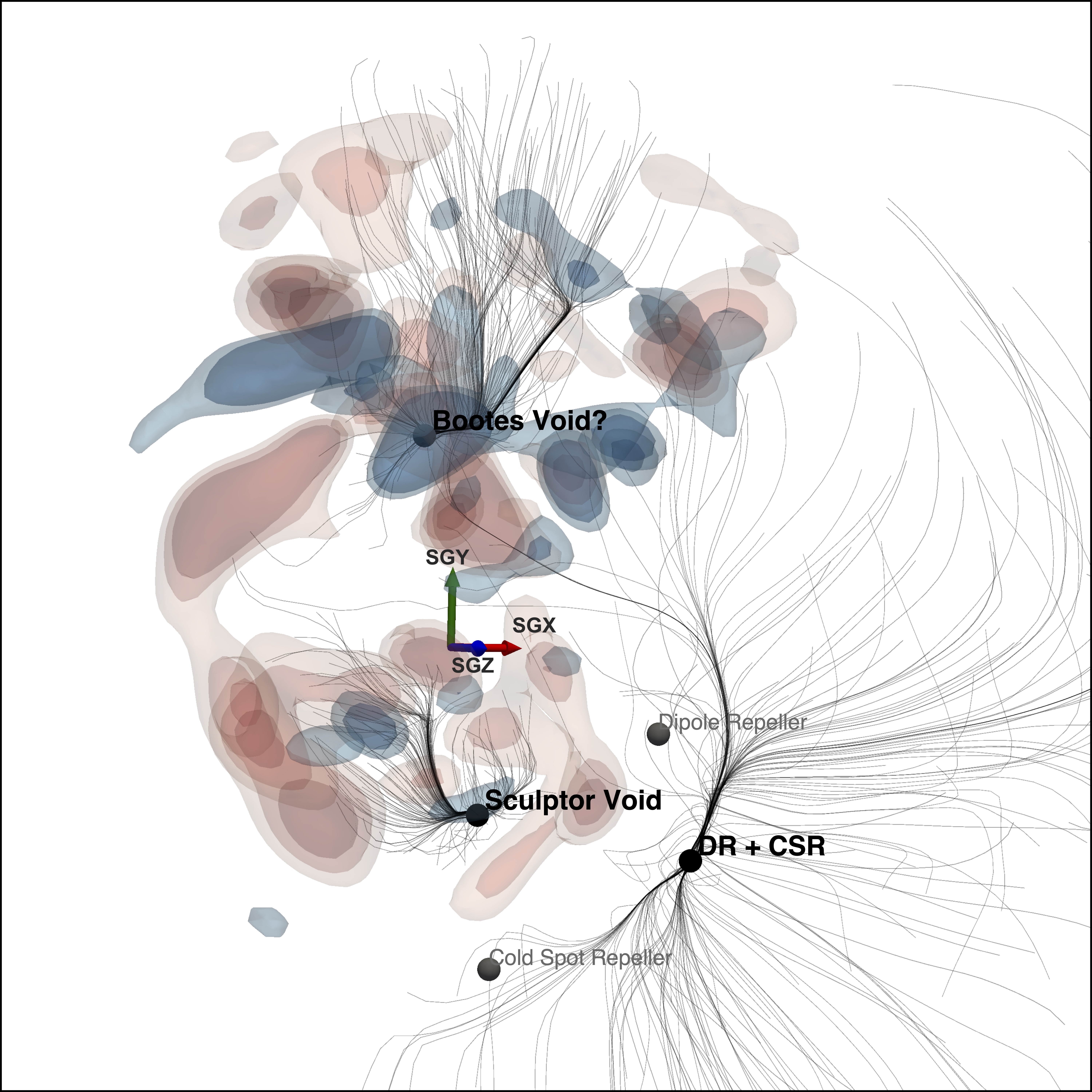}\\
\includegraphics[width=0.99\columnwidth,angle=-0]{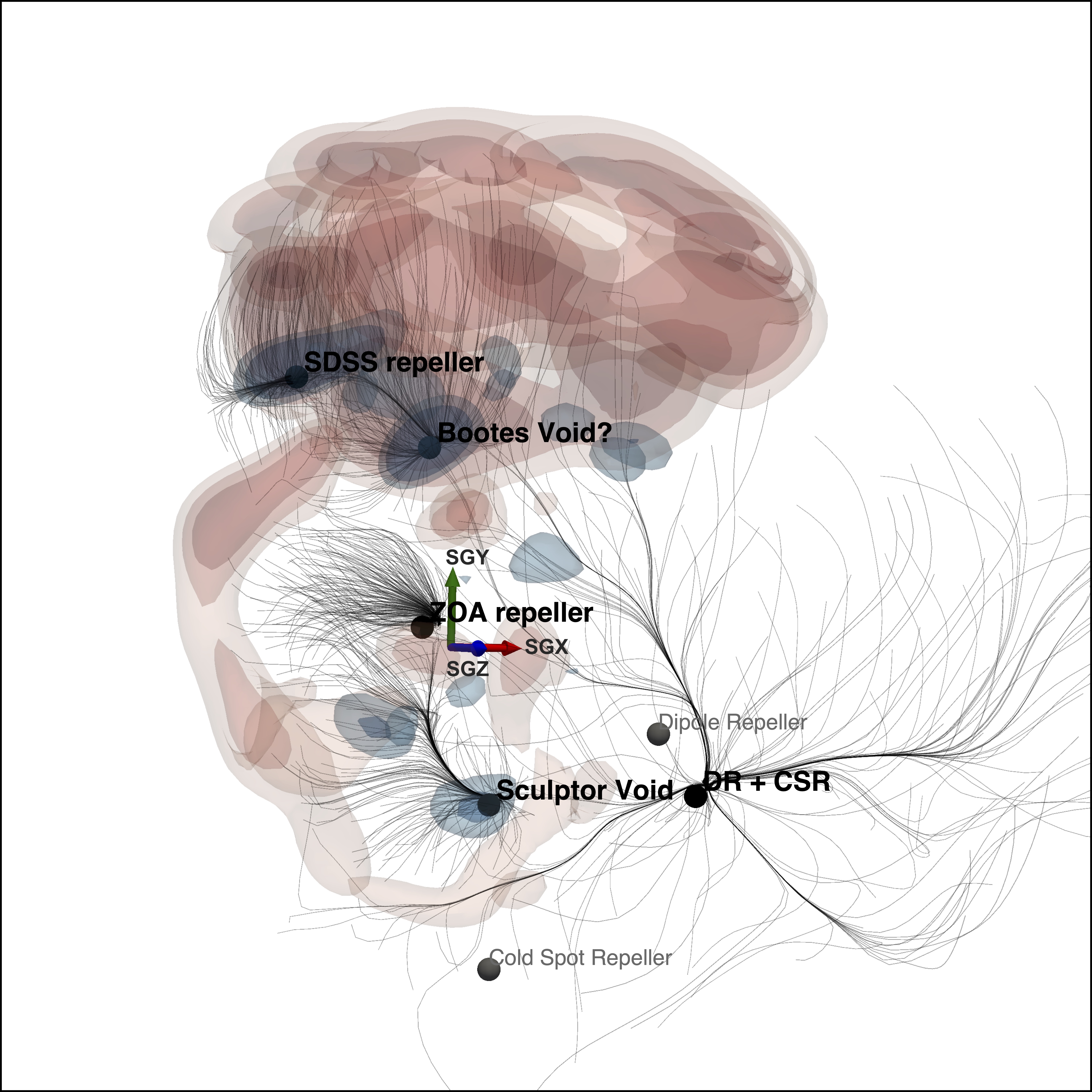}
\caption{Repellers in the local universe reconstructed from the CosmicFlows-4 galaxies (top) and groups (bottom). Overdensities are represented in different red shades by 3 levels of isosurfaces: 1, 1.4 and 1.75 for CF4 galaxies, and 0.8, 1.2 and 1.5 for CF4 groups. Underdensities are highlighted in less transparent, blue shades and represented by 2 levels of isosurfaces: -1.27 and -1.6 for CF4 galaxies, -1 and -1.35 for CF4 groups. Streamlines are derived by integrating the reconstructed peculiar velocity field in the backward direction (i.e integrating $-1 \times \vec{v}$). For the sake of clarity, extra noise has been removed by showing only streamlines part of basins of repulsion identified with the segmentation algorithm, listed in the bottom part of Table \ref{tab:basins}. Black spheres are located at the coordinates of repellers identified in the velocity field, i.e at the points where the streamlines are converging. The center of the reconstructed cube is located by the red, green and blue arrows of length 50 Mpc $h^{-1}$ and directing along the supergalactic cartesian SGX, SGY and SGZ axes respectively. Sketchfab interactive visualizations are available for both CF4 reconstructions: \href{https://sketchfab.com/3d-models/repellers-in-reconstruction-from-cf4-galaxies-82fa285df9764f00ac9fc279d618cffc}{galaxies} and \href{https://sketchfab.com/3d-models/repellers-in-reconstruction-from-cf4-groups-3d266cb55bd44ab0a62e92bd9aae1738}{groups}.}
\label{fig:Vrepellers}
\end{center}
\end{figure}

Repellers detected in both reconstructions are
shown in Figure \ref{fig:Vrepellers} on the top panel for the ungrouped reconstruction and the bottom panel for the grouped reconstruction. In order to focus on repellers, and underdensities, overdensities are only represented by the three highest levels of isosurfaces as in Figure \ref{fig:Vattractors}. Additionally, underdensities are plotted as 2 levels of isosurfaces. The red, green and blue arrows pointing towards SGX, SGY and SGZ respectively indicate the orientation of the visualization. Black spheres are placed at the coordinates of the repellers detected in each reconstruction, and the velocity field inside the associated basins of repulsion is represented by streamlines, computed in the backward direction hence converging to the locations of the repellers. The two grey spheres indicate the positions of the Dipole and Cold Spot Repeller \citep{2017NatAs...1E..36H,2017ApJ...847L...6C}.

\begin{table*}[h!]
\caption{Gravitational basins identified in the peculiar velocity field reconstructed from the CosmicFlows-4 individual galaxies (columns 2 to 8) and groups (columns 9 to 15). Column 1 lists the names of structures identified as basins of attraction (lines 3 to 15) and basins of repulsion (lines 16 to 20) in each reconstruction. Columns 2-9, 3-10 and 4-11 give the equatorial coordinates (R.A. and Dec., both in degrees) and the velocity in the CMB frame of rest $V_\mathrm{CMB}$ in km/s. Columns 5-12, 6-13 and 7-14 give the supergalactic cartesian coordinates of the corresponding attractor or repeller, SGX, SGY and SGZ, respectively, in units of Mpc $h^{-1}$. Columns 8-15 provides the volume of the gravitational basins (attraction or repulsion), in units of 10$^6$ (Mpc $h^{-1}$)$^3$. Lines are left blank if no basin corresponding to the respective structure has been identified in the reconstructions.} 
\label{tab:basins}
\begin{center}
\begin{tabular}{l|rrrrrrc|rrrrrrc}
\hline
                  & \multicolumn{7}{c}{CF4 galaxies}  & \multicolumn{7}{c}{CF4 groups} \\
                  & R.A. & Dec. & $V_\mathrm{CMB}$ & SGX  & SGY & SGZ   & Volume & R.A. & Dec. & $V_\mathrm{CMB}$ & SGX & SGY  & SGZ  & Volume \\
\hline
                 Laniakea         &  263 &   -39 &    7,412 &   -62 &   -8 &   39 &   1.9 &      &       &          &       &      &      &       \\
                 Apus             &  199 &   -75 &   13,096 &  -125 &  -23 &  -31 &   9.5 &      &       &          &       &      &      &       \\
                 Hercules         &  237 &     9 &   11,213 &   -39 &   70 &   78 &   3.1 &  234 &     7 &   12,002 &   -47 &   78 &   78 &   0.8 \\
                 Lepus            &   99 &   -49 &   13,189 &   -62 &  -39 & -109 &   8.1 &   88 &   -43 &   14,427 &   -47 &  -55 & -125 &   6.9 \\
                 Perseus-Pisces   &   56 &    21 &    6,102 &    47 &  -23 &  -31 &   4.8 &   62 &    18 &    6,536 &    47 &  -23 &  -39 &   2.0 \\
                 Shapley          &  199 &   -40 &   15,468 &  -141 &   62 &  -16 &   7.9 &  199 &   -40 &   15,468 &  -141 &   62 &  -16 &  27.0 \\
                 SDSS-1a          &  133 &    24 &   34,117 &   117 &  227 & -227 & 184.8 &  133 &    38 &   28,276 &   148 &  195 & -141 &  47.0 \\
                 SDSS-1b          &      &       &          &       &      &      &       &  129 &    14 &   29,294 &    70 &  172 & -227 &  47.1 \\
                 SDSS-1c          &      &       &          &       &      &      &       &  154 &     2 &   29,903 &   -47 &  234 & -180 &  67.2 \\
                 SDSS-1d          &      &       &          &       &      &      &       &  153 &    31 &   29,595 &    86 &  258 & -117 &  17.7 \\
                 SDSS-2a          &  209 &    12 &   32,979 &  -102 &  297 &  102 & 169.8 &  214 &    16 &   29,179 &   -70 &  258 &  117 &  60.7 \\
                 SDSS-2b          &  218 &    36 &   33,007 &    31 &  289 &  156 & 130.2 &  213 &    38 &   29,615 &    39 &  266 &  125 &  49.3 \\
                 SDSS-2c          &      &       &          &       &      &      &       &  237 &    28 &   27,632 &    -8 &  195 &  195 &  59.0 \\
\hline
                 Sculptor Void    &   18 &   -16 &   11,294 &    23 & -109 &  -16 &   1.3 &   10 &   -13 &   10,423 &    23 & -102 &    0 &   2.5 \\
                 near Bootes Void &  194 &    13 &   12,908 &   -31 &  125 &    8 &   7.8 &  198 &    13 &   12,229 &   -31 &  117 &   16 &   4.4 \\
                 SDSS repeller    &      &       &          &       &      &      &       &  194 &    -9 &   19,089 &  -109 &  156 &   -8 &   1.0 \\
                 DR + CSR         &  360 &    22 &   19,795 &   133 & -133 &   62 & 163.0 &    2 &    33 &   17,764 &   141 &  -94 &   55 &  96.2 \\
                 ZOA repeller     &      &       &          &       &      &      &       &  111 &   -12 &    5,524 &     0 &    8 &  -55 &   0.9 \\
\hline
\end{tabular}
\end{center}
\end{table*}

\end{document}